\documentclass{ws-mpla}
\usepackage[super]{cite}
\usepackage{graphics,psfrag}
\usepackage{amsmath,amssymb,epsfig,euscript,array,cite,cancel,color}
\usepackage{cases,empheq}
\usepackage[hidelinks=true]{hyperref}
\usepackage{url}
\usepackage{bbm}

\usepackage{siunitx}

\newcommand\blfootnote[1]{%
  \begingroup
  \renewcommand\thefootnote{}\footnote{#1}%
  \addtocounter{footnote}{-1}%
  \endgroup
}

\newcommand{\be}{\begin{equation}}
\newcommand{\ee}{\end{equation}}
\newcommand{\eq}[1]{(\ref{#1})}
\newcommand{\bit}{\begin{itemize}}  \newcommand{\eit}{\end{itemize}}
 
\renewcommand*{\thefootnote}{\fnsymbol{footnote}}
\def\b{\beta}         
\def\g{\gamma}

\def\k{\kappa}  
   
\def\m{\mu}   
  
\def\p{\pi}   
\def\r{\rho}

\def\cD{{\cal D}}

\def\pa{\partial}

\def\ove#1{\frac{1}{#1}}

\def\RR{\mathbb{R}}
  
\def\ZZ{\mathbb{Z}}  
  
\def\bcomment#1{}

\newcommand{\lrbrk}[1]{\left(#1\right)}

\begin{document}
\markboth{A. Tita \& P. Vanichchapongjaroen}
{Bound states of Newton's equivalent finite square well}

\catchline{}{}{}{}{}

\title{
Bound states of Newton's equivalent finite square well}

\author{\footnotesize Amornthep Tita${}^\#$ and Pichet Vanichchapongjaroen${}^{\dagger}$\footnote{Corresponding author}}

\address{The Institute for Fundamental Study ``The Tah Poe Academia Institute'',\\
Naresuan University, Phitsanulok 65000, Thailand\\
${}^\#$amornthept58@email.nu.ac.th\\
${}^\dagger$pichetv@nu.ac.th}

\maketitle

\blfootnote{Preprint of an article submitted for consideration in Modern Physics Letters A \copyright\ 2018, World Scientific Publishing Company, https://www.worldscientific.com/worldscinet/mpla.}

\pub{Received (Day Month Year)}{Revised (Day Month Year)}

\begin{abstract}
In this paper,
a 1-parameter family of Newton's equivalent Hamiltonians (NEH)
for finite square well potential is analyzed
in order to obtain bound state energy spectrum
and wavefunctions.
For a generic potential,
each of the NEH is classically equivalent to one another
and to the standard Hamiltonian yielding Newton's equations.
Quantum mechanically, however,
they are expected to be differed from each other.
The Schr\"{o}dinger's equation coming from each of NEH
with finite square well potential is an infinite
order differential equation.
The matching conditions therefore demand
the wavefunctions to be infinitely differentiable
at the well boundaries.
To handle this, we provide a way to consistently truncate
these conditions.
It turns out as expected that
bound state energy spectrum and wavefunctions
are dependent on the parameter $\b,$
which is used to characterize different NEH.
As $\b\to 0,$ the energy spectrum
coincides with that from
the standard quantum finite square well.

\end{abstract}

\ccode{PACS Nos.: 03.65.-w, 03.65.Ge}

\section{Introduction}
Degasperis and Ruijsenaars, \cite{degasperis2001newton} constructed
a one-parameter family of 
Hamiltonians whose
classical version, in conservative system, is given by
\be\label{NEH}
H_\b = \ove{\b^2 m}\lrbrk{(1+2m\b^2 V(x))^{1/2}\cosh(\b p) - 1},
\ee
where $\b$ is a real parameter.
Each Hamiltonian in this family
gives rise, via Hamilton's equation,
to the standard Newton's equation
\be
m\ddot x + V'(x) = 0.
\ee
Furthermore, in the limit $\b\to 0$
it recovers the standard Hamiltonian,
i.e.
\be
\lim_{\b\to 0} H_\b = \frac{p^2}{2m} + V(x).
\ee
Degasperis-Ruijsenaars construction 
provides one possible class of solutions
to the inverse problem \cite{Helmholtz1887},
which states that there can be many different Lagrangians
and, as a consequence, Hamiltonians
which lead the same equations of motion.
For some developments in this direction,
see for example Refs.~\refcite{currie1966q,santilli1978foundations,Morandi:1990su,henneaux1982equations,henneaux1982inverse,hojman1981inverse}.  

We will call the Hamiltonians~\eq{NEH}
and their quantum versions as Newton's equivalent Hamiltonians (NEH),
and call the quantum theory that is based on NEH
as Newton's equivalent quantum mechanics (NEQM),
while the quantum mechanics based on standard Hamiltonians
will be abbreviated simply as QM.

In order to quantize each Hamiltonian~\eq{NEH},
one inevitably faces ordering ambiguity.
But instead of directly addressing
the ambiguity by studying all the possible orderings,
we will simply choose a particular
ordering and work on it.
This is also the approach
adopted in Ref.~\refcite{degasperis2001newton},
which studied NEH of simple harmonic oscillator
by focusing on a particular operator ordering of NEH.
Then after presented their result based on this particular ordering,
they proceeded to discuss the results for some other orderings.

Although classically every Hamiltonian in the family describes
the same theory, it is not guaranteed that quantum mechanically
they would agree to each other \cite{kennedy1965note,calogero2004quantization,cislo2001extent}.
For example, the analysis of NEQM simple harmonic oscillator
by Ref.~\refcite{degasperis2001newton} showed that although the energy spectrum
(after an appropriate constant shift in the Hamiltonians)
is independent of the parameter $\b,$
the wavefunctions differ from their counterpart
in QM simple harmonic oscillator.

Apart from NEQM simple harmonic oscillator discussed by Ref.~\refcite{degasperis2001newton},
the only other NEQM system being discussed
in the literature
is an NEQM plane rotator \cite{calogero2012newton}.
One of the reasons that there only seem to be
only a couple of NEQM analyzed so far
is perhaps due to
the complication of NEH in general.
Schr\"odinger's equations for these systems
are infinite dimensional differential equations.

In this paper, we will turn our attention to NEQM finite square well.
In particular, we will analyze the bound state energy levels
and their corresponding wavefunctions.
In the QM counterpart, the strategy to analyze the system
is that one solves Schr\"odinger's equation
separately for the regions inside and outside of the well.
Then one matches the wavefunctions on the boundaries
such that they are continuous and differentiable at least once.
The reason for this requirement is thanks to the fact that
the
Schr\"odinger's equation is a second order differential equation.
For NEQM case, since the Schr\"odinger's equation is an infinite order differential equation,
the wavefunctions have to be infinitely differentiable
on the well boundaries.
The full task is therefore very complicated, if not impossible
to do so numerically.
So we will provide a way to approximate the results
and argue how the approximations are justified.

This paper is organized as follows.
In Section~\ref{sec:QMfin} we review QM finite square well system.
We emphasize on an approach which is to be applied to
the corresponding NEQM system.
In Section~\ref{sec:NEQMfin-setup} we study NEQM finite square well system
for bound state case by making use of a systematic way to truncate
infinitely many matching conditions at the well
boundaries.
The result for bound state NEQM finite square well
are discussed in Section~\ref{sec:NEQMwell}.
We obtain the discrete energy spectrum,
and explain the behavior of spectrum in this section.
Finally, we conclude our work in Section~\ref{sec:Conclude}.

\section{QM finite square well}\label{sec:QMfin}
Let us give a review on QM finite square well
by focusing on an approach which
will be extended and applied to
our analysis on NEQM finite square well
to be given in later sections.

The Schr\"odinger's equation
for QM finite square well is given by
\be
-\frac{\hbar^2}{2m}\frac{d^2}{dx^2}\psi(x)
+V(x)\psi(x) = E\psi(x),
\ee
where
\be\label{finsq-pot}
V(x)=
\begin{cases}
0; & -a<x<a\\
V_0; & x<-a\ \&\ x>a
\end{cases}
.
\ee
This separates the position space into three regions:
Region I for $x<-a,$ Region II for $-a<x<a,$
and Region III for $x>a.$
By demanding the wavefunction to vanish as $x\to\pm\infty,$
the solution to Schr\"odinger's equation at this stage is given by
\be
\psi_I(x)=A e^{\k x},\qquad
\psi_{II}(x) = B_1 \cos(kx) + B_2\sin(kx),\qquad
\psi_{III}(x)=C e^{-\k x},
\ee
where
\be
k\equiv\frac{\sqrt{2mE}}{\hbar},\qquad
\k\equiv\frac{\sqrt{2m(V_0-E)}}{\hbar},
\ee
while $\psi_I, \psi_{II},$ and $\psi_{III}$ are the restrictions
of wavefunction to Region I, II, and III,
respectively.

In order to determine the constants
$A, B_1, B_2,$ and $C,$
one considers matching conditions
at the well boundaries.
At $x=-a,$
these are $\psi_I(-a) = \psi_{II}(-a),$
and $\psi_I'(-a) = \psi_{II}'(-a).$
One obtains
\be\label{ABC-1}
A e^{-\k a}
=B_1 \cos(ka) - B_2\sin(ka),
\ee
\be\label{ABC-2}
A\k e^{-\k a}
=B_1 k \sin(ka) + B_2 k\cos(ka).
\ee
At $x=a$
the matching conditions are
$\psi_{III}(a) = \psi_{II}(a),$
and $\psi_{III}'(a) = \psi_{II}'(a).$
These give
\be\label{ABC-3}
C e^{-\k a}
=B_1 \cos(ka) + B_2\sin(ka),
\ee
\be\label{ABC-4}
-C\k e^{-\k a}
=-B_1 k \sin(ka) + B_2 k\cos(ka).
\ee
Equations~\eq{ABC-1}-\eq{ABC-4}
can be rewritten in matrix form
\be\label{zeromode-eqn}
M
\begin{pmatrix}
A\\
B_1\\
B_2\\
C
\end{pmatrix}
=
\begin{pmatrix}
0\\
0\\
0\\
0
\end{pmatrix}
,
\ee
where
\be
M
=
\begin{pmatrix}
e^{-\k a} & -\cos(ka) & \sin(ka) & 0\\
\k e^{-\k a}  & -k\sin(ka) & -k\cos(ka) & 0\\
0 & -\cos(ka) & -\sin(ka) & e^{-\k a}\\
0 & k\sin(ka) & -k\cos(ka) & -\k e^{-\k a}
\end{pmatrix}
.
\ee
In order for Eq.~\eq{zeromode-eqn}
to have a non-trivial solution,
we need $\det M = 0.$
This implies
\be
2(\k\cos(k a)-k\sin(ka))(\k\sin(k a)+k\cos(ka)) = 0,
\ee
which gives
\be\label{kappak}
\k = k\tan(k a),\qquad
\textrm{or}\qquad
\k = -k\cot(ka).
\ee
In order to obtain energy spectrum for bound states,
these conditions are solved together
with
\be\label{kappakV0}
\k^2 + k^2 = \frac{2m V_0}{\hbar^2}
\ee
for each given $V_0.$
The solving technique is standard
and can be found in most textbooks
on quantum mechanics.

Before we proceed to the case of
NEQM finite square well, let us note that
wavefunctions are either even or odd functions.
In the case of even functions, $B_2 = 0, C=A,$
and hence
\be
M^{(e)}
\begin{pmatrix}
A\\
B_1
\end{pmatrix}
=
\begin{pmatrix}
0\\
0
\end{pmatrix}
\qquad\textrm{with}\;\;\;
M^{(e)}
=
\begin{pmatrix}
e^{-\k a} & -\cos(ka)\\
\k e^{-\k a}  & -k\sin(ka)
\end{pmatrix}
.
\ee
For the case of odd functions, $B_1 = 0, C= -A,$
so
\be
M^{(o)}
\begin{pmatrix}
A\\
B_2
\end{pmatrix}
=
\begin{pmatrix}
0\\
0
\end{pmatrix}
\qquad\textrm{with}\;\;\;
M^{(o)}
=
\begin{pmatrix}
e^{-\k a} & \sin(ka)\\
\k e^{-\k a}  & -k\cos(ka)
\end{pmatrix}
.
\ee
By using appropriate row and column operations,
the matrix $M$ can essentially be factorized as
\be
M
=
\begin{pmatrix}
 1 & 0 & -\frac{1}{2} & 0 \\
 0 & 1 & 0 & -\frac{1}{2} \\
 1 & 0 & \frac{1}{2} & 0 \\
 0 & -1 & 0 & -\frac{1}{2}
\end{pmatrix}
\begin{pmatrix}
M^{(e)} & 0\\
0 & M^{(o)}
\end{pmatrix}
\begin{pmatrix}
 \frac{1}{2} & 0 & 0 & \frac{1}{2} \\
 0 & 1 & 0 & 0 \\
 -1 & 0 & 0 & 1 \\
 0 & 0 & -2 & 0
\end{pmatrix}
,
\ee
and hence its determinant is factorized as
\be
\det M = -2\det M^{(e)}\det M^{(o)}.
\ee
This factorization confirms that
wavefunctions can only be either even
or odd functions.

\section{Set up for finite square well in NEQM}\label{sec:NEQMfin-setup}
\subsection{Schr\"odinger's equation and matching conditions}
Let us study NEQM finite square well,
whose Hamiltonian is given by, \cite{degasperis2001newton}
\be\label{NEQMfswHam}
H
=\ove{2\b^2 m}\lrbrk{(1+i\b \sqrt{2m V(x)})^{1/2}e^{-i\hbar\b\pa_x}(1-i\b \sqrt{2m V(x)})^{1/2} + (i\to -i)}-\ove{\b^2 m},
\ee
where $V(x)$ is given by Eq.~\eq{finsq-pot}
This potential
is constant in each region.
So by restricting to any particular region, we may write
$V(x)=V_c.$
Then it is clear that $V_c$ commutes with $\pa_x,$
and so can be moved to the leftmost.
That is, 
the Schr\"odinger's equation in each region can be written in the form
\be
\lrbrk{\ove{2\b^2 m}(1+\b^2 2m V_c)^{1/2}\lrbrk{e^{-i\hbar\b\pa_x} + e^{i\hbar\b\pa_x}}-\ove{\b^2 m}}\psi(x) = E\psi(x).
\ee
Imposing an ansatz $\psi(x) = e^{(\g_r + i\g_i) x},$
where $\g_r,\g_i\in\RR,$
it can be seen that the values of $\g_r$ and $\g_i$
fall into one of the three cases, sorted by the range
of validity for energy:

\textbf{Case $1$:}
\be
\g_r = \frac{\p (2n+1)}{\hbar\b},\qquad
\g_i = \ove{\hbar\b}\log\lrbrk{-\frac{1+m\b^2 E}{\sqrt{1+2m\b^2 V_c}}\pm\sqrt{\frac{(1+m\b^2 E)^2}{1+2m\b^2 V_c}-1}},
\ee
for $n\in\ZZ.$
This case is valid for
\be
E\leq \frac{-\sqrt{1+2m\b^2 V_c}-1}{m\b^2}.
\ee

\textbf{Case $2$:}
\be
\cos(\hbar\b\g_r) = \frac{1+m\b^2 E}{\sqrt{1+2m\b^2 V_c}},\qquad
\g_i = 0.
\ee
This case is valid for
\be
\frac{-\sqrt{1+2m\b^2 V_c}-1}{m\b^2} \leq E \leq \frac{\sqrt{1+2m\b^2 V_c}-1}{m\b^2}.
\ee

\textbf{Case $3$:}
\be
\g_r = \frac{2n\p}{\hbar\b},\qquad
\g_i = \ove{\hbar\b}\log\lrbrk{\frac{1+m\b^2 E}{\sqrt{1+2m\b^2 V_c}}\pm\sqrt{\frac{(1+m\b^2 E)^2}{1+2m\b^2 V_c}-1}},
\ee
for $n\in\ZZ.$
This case is valid for
\be
E\geq \frac{\sqrt{1+2m\b^2 V_c}-1}{m\b^2}.
\ee

In the QM limit $\b\to 0,$
case $2$ applies to regions I and III,
whereas case $3$ applies to region II.
So let us assume that this is also the case for
other values of $\b.$
Therefore, the discrete energy spectrum should fall within the range
\be
0\leq E\leq \frac{\sqrt{1+2m\b^2 V_0}-1}{m\b^2}.
\ee
The wavefunctions which satisfy regularity requirement, $\psi(x)\to 0$ as $x\to\pm\infty,$
are given by
\be\label{psiI}
\psi_I(x) = \sum_{l=0}^\infty A_{1,l}\exp\lrbrk{\frac{2\p l x}{\hbar\b}}\exp\lrbrk{\r x}
+\sum_{l=1}^\infty A_{-1,l}\exp\lrbrk{\frac{2\p l x}{\hbar\b}}\exp\lrbrk{-\r x},
\ee
\be\label{psiII}
\psi_{II}(x) = \sum_{l=-\infty}^\infty B_{a,l}\exp\lrbrk{\frac{2\p l x}{\hbar\b}}\cos(\m x)
+\sum_{l=-\infty}^\infty B_{b,l}\exp\lrbrk{\frac{2\p l x}{\hbar\b}}\sin(\m x),
\ee
\be\label{psiIII}
\psi_{III}(x) = \sum_{l=1}^\infty C_{1,l}\exp\lrbrk{\frac{-2\p l x}{\hbar\b}}\exp\lrbrk{\r x}
+\sum_{l=0}^\infty C_{-1,l}\exp\lrbrk{\frac{-2\p l x}{\hbar\b}}\exp\lrbrk{-\r x},
\ee
where
\be
\r \equiv \ove{\hbar\b}\cos^{-1}\lrbrk{\frac{1+m\b^2 E}{\sqrt{1+2m\b^2 V_0}}},
\ee
\be
\m\equiv \ove{\hbar\b}\log\lrbrk{1+m\b^2 E + \sqrt{(1+m\b^2 E)^2-1}}.
\ee
The coefficients will be determined
by the matching conditions at $x=\pm a.$
Since the Schr\"odinger's equation~\eq{NEQMfswHam}
is a differential equation of infinite order,
the wavefunction has to be differentiable
infinitely many times at the boundaries $x=\pm a$ of the well.

The matching conditions $\psi_I^{(n)}(-a) = \psi_{II}^{(n)}(-a),
\psi_{III}^{(n)}(a) = \psi_{II}^{(n)}(a)$ for all $n = 0,1,2,\cdots,$
can be translated to a matrix equation of the form
\be
M_{\infty}
\begin{pmatrix}
A_{\cdots}\\
\vdots\\
B_{\cdots}\\
\vdots\\
C_{\cdots}\\
\vdots
\end{pmatrix}
=
\begin{pmatrix}
0\\
\vdots\\
0\\
\vdots\\
0\\
\vdots
\end{pmatrix}
.
\ee
In order for the wavefunction to be nontrivial,
the matrix $M_{\infty}$ has to be singular:
\be\label{detMinfty0}
\det M_{\infty} = 0.
\ee
The roots of this condition give
the energy spectrum $E$ for any fixed $\b$ and $V_0.$
In practice, computation of determinant of $M_{\infty}$
would require the matrix to be truncated,
say to an $N\times N$ matrix, called $M_N.$
As $N\to\infty,$ one would expect that the roots of
\be
\det M_N = 0
\ee
would converge to those of Eq.~\eq{detMinfty0}.

\subsection{A systematic truncation}
Instead of arbitrarily cutting off rows and columns of $M_{\infty},$
let us propose a systematic way to truncate $M_{\infty}.$
We will start from truncating the wavefunction Eqs.~\eq{psiI}-\eq{psiIII} to
\be\label{psiI-trunc}
\psi_I(x) = \sum_{l=0}^L A_{1,l}\exp\lrbrk{\frac{2\p l x}{\hbar\b}}\exp\lrbrk{\r x}
+\sum_{l=1}^L A_{-1,l}\exp\lrbrk{\frac{2\p l x}{\hbar\b}}\exp\lrbrk{-\r x},
\ee
\be
\psi_{II}(x) = \sum_{l=-L}^L B_{a,l}\exp\lrbrk{\frac{2\p l x}{\hbar\b}}\cos(\m x)
+\sum_{l=-L}^L B_{b,l}\exp\lrbrk{\frac{2\p l x}{\hbar\b}}\sin(\m x),
\ee
\be\label{psiIII-trunc}
\psi_{III}(x) = \sum_{l=1}^L C_{1,l}\exp\lrbrk{\frac{-2\p l x}{\hbar\b}}\exp\lrbrk{\r x}
+\sum_{l=0}^L C_{-1,l}\exp\lrbrk{\frac{-2\p l x}{\hbar\b}}\exp\lrbrk{-\r x},
\ee
for $L\in\{0,1,2,\cdots\}.$ So there are $4(2L+1)$ arbitrary constants. Furthermore,
we require matching conditions to be continuous up to order $4L+1$ derivative.
That is we demand
\be\label{contdcond-trunc}
\psi_I^{(n)}(-a) = \psi_{II}^{(n)}(-a),\qquad
\psi_{III}^{(n)}(a) = \psi_{II}^{(n)}(a)
\ee
for all $n = 0,1,2,\cdots,4L+1.$
This gives rise to a matrix $M_{4(2L+1)}.$
Then for each
$\b$ and $V_0,$
we solve for the roots of $\det M_{4(2L+1)} = 0,$
and study how they
would converge
as $L$ increases.

In fact,
$\det M_{4(2L+1)}$ can be factorized
as a product of determinants of matrices
corresponding to even and odd wavefunctions.
To show this, let us rewrite the wavefunction~\eq{psiI-trunc}-\eq{psiIII-trunc}
as, note the renaming of the arbitrary constants,
\be\label{psiI-trunc-new}
\psi_I(x) = P(x)\vec{A},
\ee
\be
\psi_{II}(x) = Q_e(x)\vec{B}_e + Q_o(x)\vec{B}_o,
\ee
\be\label{psiIII-trunc-new}
\psi_{III}(x) = P(-x)\vec{C},
\ee
where $P(x), Q_e(x), Q_o(x)$ are row vectors:
\be
P(x) =
\begin{pmatrix}
e^{\r x},
& e^{\frac{2\p x}{\hbar\b}+\r x},
& \cdots,
& e^{\frac{2\p L x}{\hbar\b}+\r x},
& e^{\frac{2\p x}{\hbar\b}-\r x},
& \cdots,
& e^{\frac{2\p L x}{\hbar\b}-\r x}
\end{pmatrix}
,
\ee
\be
\begin{split}
Q_e(x) &=
\bigg(
\begin{matrix}
\cos\m x,
& \cosh\lrbrk{\frac{2\p x}{\hbar\b}}\cos(\m x),
& \cdots,
& \cosh\lrbrk{\frac{2\p L x}{\hbar\b}}\cos(\m x),
\end{matrix}
\\
&\qquad\qquad
\begin{matrix}
& \sinh\lrbrk{\frac{2\p x}{\hbar\b}}\sin(\m x),
& \cdots,
& \sinh\lrbrk{\frac{2\p L x}{\hbar\b}}\sin(\m x)
\end{matrix}
\bigg)
,
\end{split}
\ee
\be
\begin{split}
Q_o(x) &=
\bigg(
\begin{matrix}
\sin\m x,
& \cosh\lrbrk{\frac{2\p x}{\hbar\b}}\sin(\m x),
& \cdots,
& \cosh\lrbrk{\frac{2\p L x}{\hbar\b}}\sin(\m x),
\end{matrix}
\\
&\qquad\qquad
\begin{matrix}
& -\sinh\lrbrk{\frac{2\p x}{\hbar\b}}\cos(\m x),
& \cdots,
& -\sinh\lrbrk{\frac{2\p L x}{\hbar\b}}\cos(\m x)
\end{matrix}
\bigg)
,
\end{split}
\ee
whereas $\vec{A}, \vec{B}_e, \vec{B}_o, \vec{C}$
are column vectors
\be
\vec{A}
=
\begin{pmatrix}
A_0\\
\vdots\\
A_{2L}
\end{pmatrix}
,
\qquad
\vec{B}_e
=
\begin{pmatrix}
B_{e,0}\\
\vdots\\
B_{e,2L}
\end{pmatrix}
,
\qquad
\vec{B}_o
=
\begin{pmatrix}
B_{o,0}\\
\vdots\\
B_{o,2L}
\end{pmatrix}
,
\qquad
\vec{C}
=
\begin{pmatrix}
C_{0}\\
\vdots\\
C_{2L}
\end{pmatrix}
.
\ee
The matching conditions then give
\be\label{contdmat}
M_{4(2L+1)}
\begin{pmatrix}
\vec{A}\\
\vec{B}_e\\
\vec{B}_o\\
\vec{C}
\end{pmatrix}
=
\begin{pmatrix}
\vec{0}\\
\vec{0}\\
\vec{0}\\
\vec{0}
\end{pmatrix}
,
\ee
where
\be
M_{4(2L+1)}
=
\begin{pmatrix}
\cD P(-a) & -\cD Q_e(-a) & -\cD Q_o(-a) & 0\\
0 & -\cD^* Q_e(-a) & \cD^* Q_o(-a) & \cD^* P(-a)
\end{pmatrix}
.
\ee
Here we defined the operators $\cD, \cD^*$
such that
\be
\cD P(x)
\equiv
\begin{pmatrix}
P(x)\\
P'(x)\\
P''(x)\\
P'''(x)\\
\vdots\\
P^{(4L+1)}(x)
\end{pmatrix}
,
\qquad
\cD^*P(x)
\equiv
\begin{pmatrix}
P(x)\\
-P'(x)\\
P''(x)\\
-P'''(x)\\
\vdots\\
-P^{(4L+1)}(x)
\end{pmatrix}
,
\ee
and similarly for $\cD Q_e, \cD^*Q_e, \cD Q_o, \cD^*Q_o.$
We have also used the identities
\be
\cD Q_e(-x) = \cD^*Q_e(x),\qquad
\cD Q_o(-x) = -\cD^*Q_o(x).
\ee
Choosing even wavefunctions corresponds to
setting $\vec{B}_o = \vec{0}, \vec{C} = \vec{A}.$
So Eq.~\eq{contdmat} becomes,
after removing redundant rows,
\be
M^{(e)}_{4(2L+1)}
\begin{pmatrix}
\vec{A}\\
\vec{B}_e
\end{pmatrix}
=
\vec{0}
,\qquad
\textrm{with}\quad
M^{(e)}_{4(2L+1)}
\equiv
\begin{pmatrix}
\cD P(-a), &- \cD Q_e(-a)
\end{pmatrix}
\ee
Similarly,
choosing odd wavefunction corresponds to
setting $\vec{B}_e = \vec{0}, \vec{C} = -\vec{A},$
and Eq.~\eq{contdmat} reduces to
\be
M^{(o)}_{4(2L+1)}
\begin{pmatrix}
\vec{A}\\
\vec{B}_o
\end{pmatrix}
=
\vec{0}
,
\qquad\textrm{with}\quad
M^{(o)}_{4(2L+1)}
\equiv
\begin{pmatrix}
\cD P(-a), &- \cD Q_o(-a)
\end{pmatrix}
.
\ee
After performing row and column operations,
it can easily be seen that
\be
\det M_{4(2L+1)}
=
-2\det M^{(e)}_{4(2L+1)} \det M^{(o)}_{4(2L+1)}
.
\ee

\section{Results for finite square well in NEQM}\label{sec:NEQMwell}
Let us now discuss how to obtain
bound state energy spectrum for finite square well in NEQM.
For a fixed $\b$ and $V_0,$
we construct truncated matrices $M^{(e)}_{4(2L+1)}$
and $M^{(o)}_{4(2L+1)}$
by using matching conditions
for even and odd wavefunctions respectively.
We substitute
\be
\r = \ove{\hbar\b}\cos^{-1}\lrbrk{\frac{\cosh(\hbar\b\m)}{\sqrt{1+2m\b^2 V_0}}}
\ee
into these matrices then solve for the roots $\m$ of
$\det M^{(e)}_{4(2L+1)} = 0$ and of $\det M^{(o)}_{4(2L+1)} = 0.$
They can then be used to obtained eigenstate energy
from
\be
E = \frac{\cosh(\hbar\b\m)-1}{m\b^2}.
\ee

In order to do so numerically,
we proceed as follows.
For each $\b$ and $V_0,$
two of the $2(2L+1)\times 2(2L+1)$ matrices
and their determinants have to be computed
as functions of $\m.$
Then we make equal-spaced samplings
of $\m$ in the range
\be
0\leq \m\leq \ove{\hbar\b}\log\lrbrk{\sqrt{1+2m\b^2 V_0} + \sqrt{2m\b^2 V_0}},
\ee
and interpolate to see where $\det M^{(e)}_{4(2L+1)}$
or $\det M^{(o)}_{4(2L+1)}$ cross the horizontal axis.
Then use these values of $\m$ to obtain energy $E.$

\begin{table}[h]
	\tbl{Ground state energy ($ma^2 E/\hbar$)
		for $L=0,1,\cdots,8$
		with $V_0 = 5000\hbar^2/(ma^2).$
		The results are obtained using $500$ equally-spaced sample values
		of $\m.$}
	{\begin{tabular}{@{}c S[table-format=1.5] S[table-format=1.5e+2] S[table-format=1.5e+2] S[table-format=1.5e+2] S[table-format=1.5e+2]@{}}
		\toprule
		&\multicolumn{5}{c}{$\b\hbar/a$}\\\cline{2-6}
$L$	&\multicolumn{1}{c}{$0.01$} &\multicolumn{1}{c}{$1.01$} & \multicolumn{1}{c}{$5.01$} & \multicolumn{1}{c}{$21.01$}& \multicolumn{1}{c}{$50.01$}\\\colrule
 0 & 1.20290 & 5.39848e-1 & 2.45167e-1 & 2.93095e-1 & 5.74541e-1 \\
 1 & 1.20028 & 5.18674e-1 & 1.77726e-1 & 6.87060e-2 & 4.47974e-2 \\
 2 & 1.19777 & 5.16123e-1 & 1.69832e-1 & 5.45742e-2 & 2.93317e-2 \\
 3 & 1.19528 & 5.15333e-1 & 1.67528e-1 & 5.02396e-2 & 2.50568e-2 \\
 4 & 1.19280 & 5.14983e-1 & 1.66555e-1 & 4.83072e-2 & 2.31343e-2 \\
 5 & 1.19035 & 5.14789e-1 & 1.66043e-1 & 4.72795e-2 & 2.20836e-2 \\
 6 & 1.18792 & 5.14665e-1 & 1.65747e-1 & 4.66899e-2 & 2.14259e-2 \\
 7 & 1.18553 & 5.14579e-1 & 1.65559e-1 & 4.62771e-2 & 2.09944e-2 \\
 8 & 1.18374 & 5.14615e-1 & 1.65131e-1 & 4.60012e-2 & 2.07552e-2\\
		\botrule
	\end{tabular}
	\label{tab-sample-gnd}}
\end{table}

As $L$ increases, the result suggests that
for each fixed $\b$ and $V_0,$
the energy for each state tends to converge.
We demonstrate this in Tables~\ref{tab-sample-gnd}-\ref{tab-sample-fst},
which show,
respectively for the ground and the first excited states,
energies for
example value of $V_0 = 5000\hbar^2/(ma^2)$
with various $\b$ and $L.$
It can be worked out from the tables that
for each fixed $\b,$ when $L$ increases
the percentage difference between two consecutive values
tends to decrease as $L$ increases.
This is except for some cases at $L=8,$
which is probably due to some numerical difficulties.
By excluding these exceptional cases,
the trend suggests
that the value of energy for each fixed $\b$ and $V_0$
should converge as $L$ increases.
Furthermore for each given state,
as $\b$ is smaller, the convergence rate is quicker.
For example, in Table~\ref{tab-sample-gnd}
using $L=3$ is already sufficient
to obtain the ground state energy for $\b = 1.01a/\hbar, V_0 = 5000\hbar^2/(ma^2),$
up to three significant figures.
However, to obtain the same precision
when $\b = 5.01a/\hbar,$ one needs $L=8$ or beyond.

\begin{table}[h]
	\tbl{First excited state energy ($ma^2 E/\hbar$)
		for $L=0,1,\cdots,8$
		with $V_0 = 5000\hbar^2/(ma^2).$
		The results are obtained using $500$ equally-spaced sample values
				of $\m.$ Each blank entry indicates that there is no data.}
	{\begin{tabular}{@{}c S[table-format=1.5] S[table-format=1.5] S[table-format=1.5e+1] S[table-format=1.5e+1] S[table-format=1.5e+1]@{}}
		\toprule
		&\multicolumn{5}{c}{$\b\hbar/a$}\\\cline{2-6}
$L$	&\multicolumn{1}{c}{$0.01$} &\multicolumn{1}{c}{$1.01$} & \multicolumn{1}{c}{$3.01$} & \multicolumn{1}{c}{$5.01$}& \multicolumn{1}{c}{$7.01$}\\\colrule
 0 & 4.81188 & 3.49359 & 1.19548e1 &  &  \\
 1 & 4.81122 & 3.17995 & 5.57617 & 1.10575e1 &  \\
 2 & 4.81098 & 3.13804 & 4.83235 & 7.95547 & 1.12967e1 \\
 3 & 4.81080 & 3.12516 & 4.61457 & 7.07036 & 9.61443 \\
 4 & 4.81062 & 3.11964 & 4.52332 & 6.70500 & 8.86594 \\
 5 & 4.81046 & 3.11676 & 4.47640 & 6.51993 & 8.47866 \\
 6 & 4.81030 & 3.11504 & 4.44886 & 6.41397 & 8.25470 \\
 7 & 4.81015 & 3.11391 & 4.43155 & 6.34692 & 8.11346 \\
 8 & 4.81008 & 3.11228 & 4.41955 & 6.30671 & 8.02452\\
		\botrule
	\end{tabular}
	\label{tab-sample-fst}}
\end{table}

Furthermore, there is a generic trend
for the first excited state and above
that in some case these states may falsely disappear
for lower values of $L,$
but in fact these states exist
as pointed out by the result for higher values of $L.$
This is demonstrated in the last two columns
of Table~\ref{tab-sample-fst}.

Note that setting $L$ too high makes the size of
the corresponding matrices too large,
thus consuming more time in computing determinants.
Furthermore, the matrices $M^{(e)}_{4(2L+1)}, M^{(o)}_{4(2L+1)}$ and their determinants
need to be computed for each sample value of $\m.$
So in the data collection for the following results,
we do not set
the truncation parameter as high as $L = 8,$
nor set the number of sample values of $\m$
as high as $500.$
We will set these numbers in such a way that
the computational time is tremendously reduced,
and at the same time considerable precision is preserved.

\begin{figure}[h]
\centering
\includegraphics[width=0.48\linewidth]{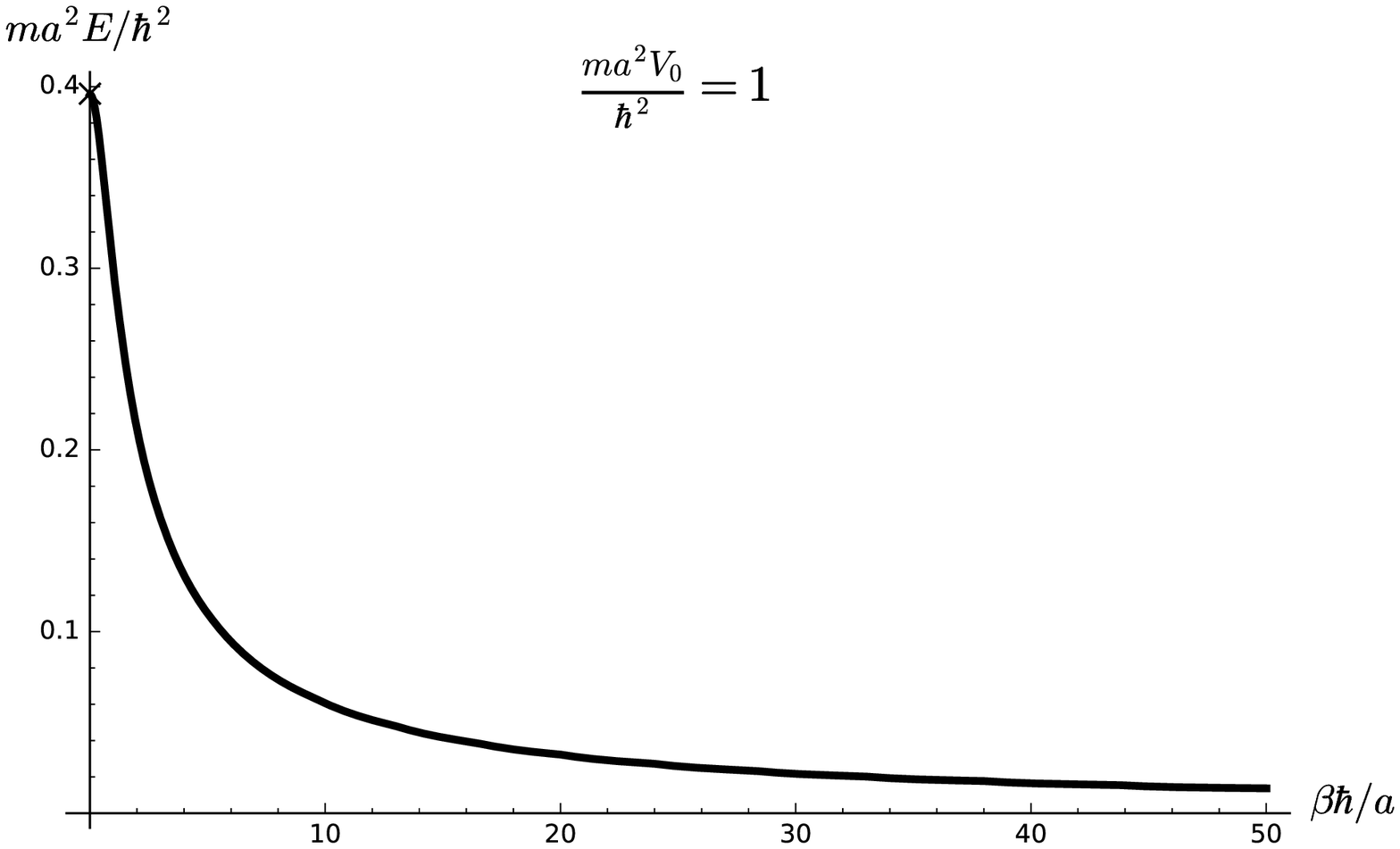}
\includegraphics[width=0.48\linewidth]{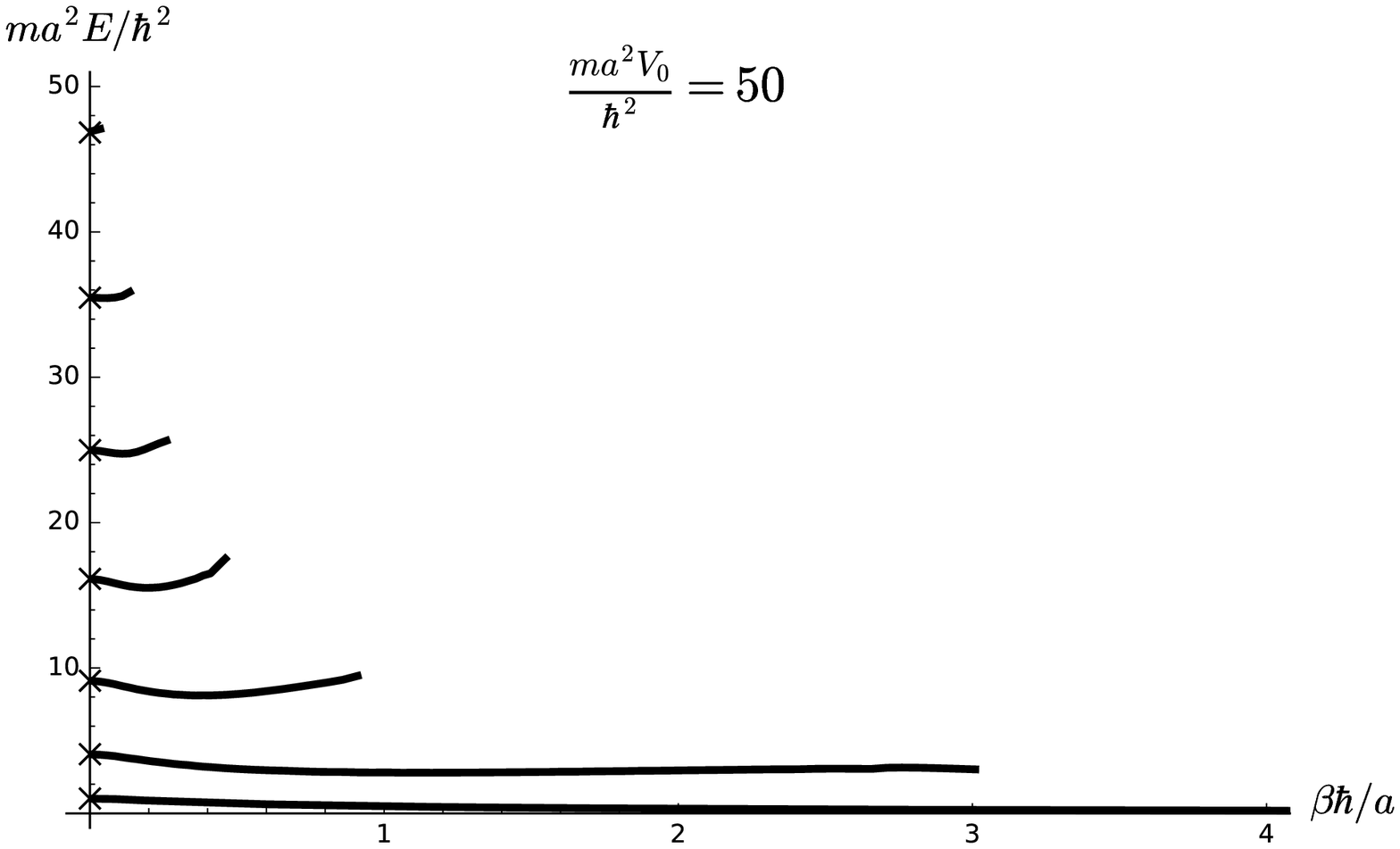}
\includegraphics[width=0.48\linewidth]{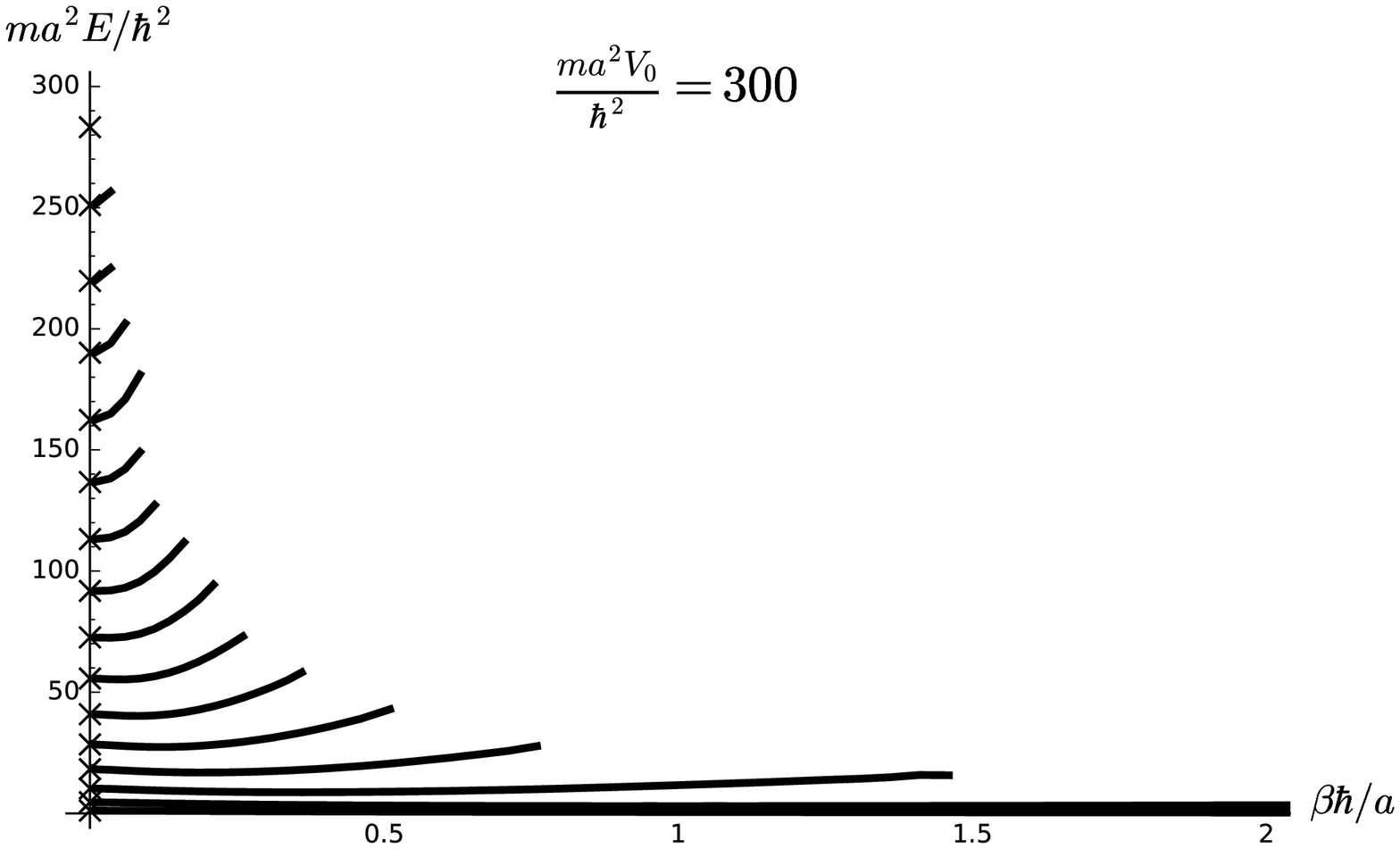}
\includegraphics[width=0.48\linewidth]{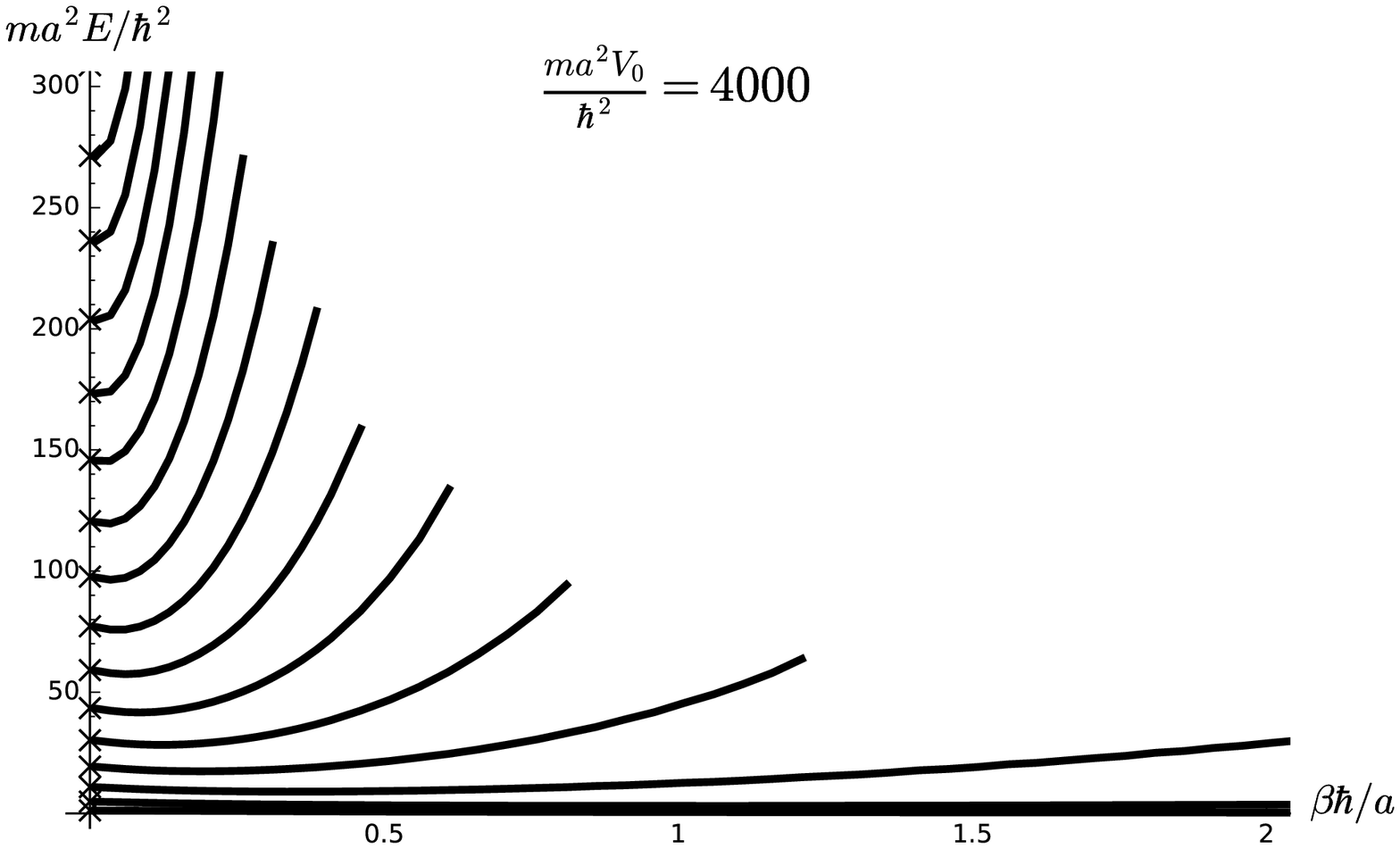}
\caption{Energy level versus $\b$ for each fixed $V_0.$
From left to right and top to bottom, $ma^2V_0/\hbar^2 = 1,50,300,4000.$
In the data collection, we set $L=6$
and use at least $50$ sampling values of $\m.$
The crosses show the energy spectrum
for QM finite square well potential.}
\label{neqm-fixv}
\end{figure}

In Figure~\ref{neqm-fixv},
we show for each fixed $V_0,$
the bound state energy spectrum $E$
as a function of $\b.$
A common feature is that as $\b$ increases,
states started to disappear one by one.
However, it seems that ground state still exists
for large values of $\b.$
We have checked that
ground state exists even at $\b = 10^6 a/\hbar, V_0 = \hbar^2/(ma^2).$
The disappearance of the energy levels for states other than the ground state
is due to the fact that
their values grow above the upper bound $(\sqrt{1+2m\b^2 V_0}-1)/(m\b^2).$
This bound has an analogy in QM finite square well,
whose bound is $V_0,$
i.e. the value of $E$ cannot be larger than $V_0.$
Note that consistently, $(\sqrt{1+2m\b^2 V_0}-1)/(m\b^2)\to V_0$
as $\b\to 0.$

We also demonstrated in the figure that
the spectrum for small $\b$
converges to the spectrum
for QM finite square well.
This convergence justifies the 
$\b\to 0$ limit. Although the result is as expected,
it is not so obvious because when $L>0,$
the wavefunction Eqs.~\eq{psiI-trunc}-\eq{psiIII-trunc}
contains expressions $\sim\exp(1/\b),$ which is singular as $\b\to 0.$
For the case of $L=0,$
due to the absence of unpleasant expressions like $\sim\exp(1/\b),$
it is easy to see that for NEQM finite square well,
the limit $\b\to 0$ indeed reduces to QM finite square well.
As for $L>0$, in order to understand these cases
let us investigate the cases $L=1,2$
where symbolic manipulation is still tractable.
We consider asymptotic expansion for $\b\to 0,$
in which $\cosh(2\p na/(\hbar\b))\sim\exp(2\p na/(\hbar\b)),
\sinh(2\p na/(\hbar\b))\sim\exp(2\p na/(\hbar\b)),$
and $\r\sim\k, \m\sim k.$
The leading order of the asymptotic expansion
for determinant of $M^{(e)}$ are given by
\be
\det M^{(e)}_{12}
\sim\lrbrk{\frac{2\p}{\b\hbar}}^{12}2^5 e^{-\k a}k\k (\k\cos(ak)-k\sin(ak)),
\ee
\be
\det M^{(e)}_{20}
\sim\lrbrk{\frac{2\p}{\b\hbar}}^{40}2^{22}3^8 e^{-\k a}k^2\k^2 (\k\cos(ak)-k\sin(ak)).
\ee
By using the relationship
$M^{(e)}\to M^{(o)}$ under the transformation
$\cos(a\m)\to\sin(a\m), \sin(a\m)\to-\cos(a\m),$
one can simply read off asymptotic expansion of
$\det M^{(o)}_{12}, \det M^{(o)}_{20}$
by simply applying the same transformation
to $\det M^{(e)}_{12}, \det M^{(e)}_{20},$ respectively.
Recall that in QM finite square well,
the energy levels associated to even (resp. odd) wavefunctions
can be obtained from the roots of Eq.~\eq{kappak}
with $\k$ given by Eq.~\eq{kappakV0},
and that $k=0$ gives a trivial wavefunction $\psi(x) = 0.$
So it can be concluded that
asymptotically the roots of $\det M_{4(2L+1)} = 0$
for $L=1,2$ coincide with the ones from QM finite square well.
We expect that the similar conclusions should also hold
for $L>2.$

\begin{figure}[h]
\centering
\includegraphics[width=0.48\linewidth]{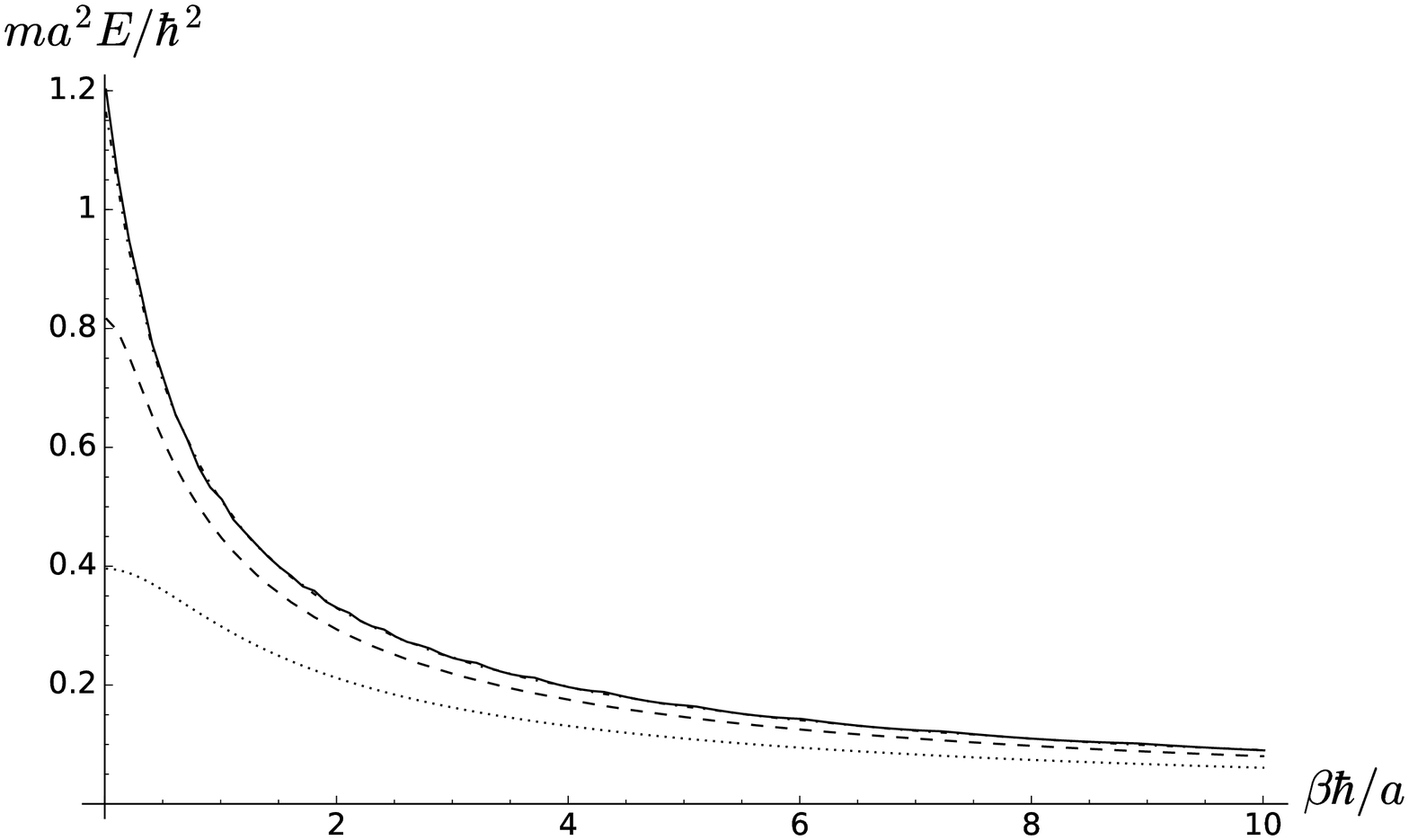}
\includegraphics[width=0.48\linewidth]{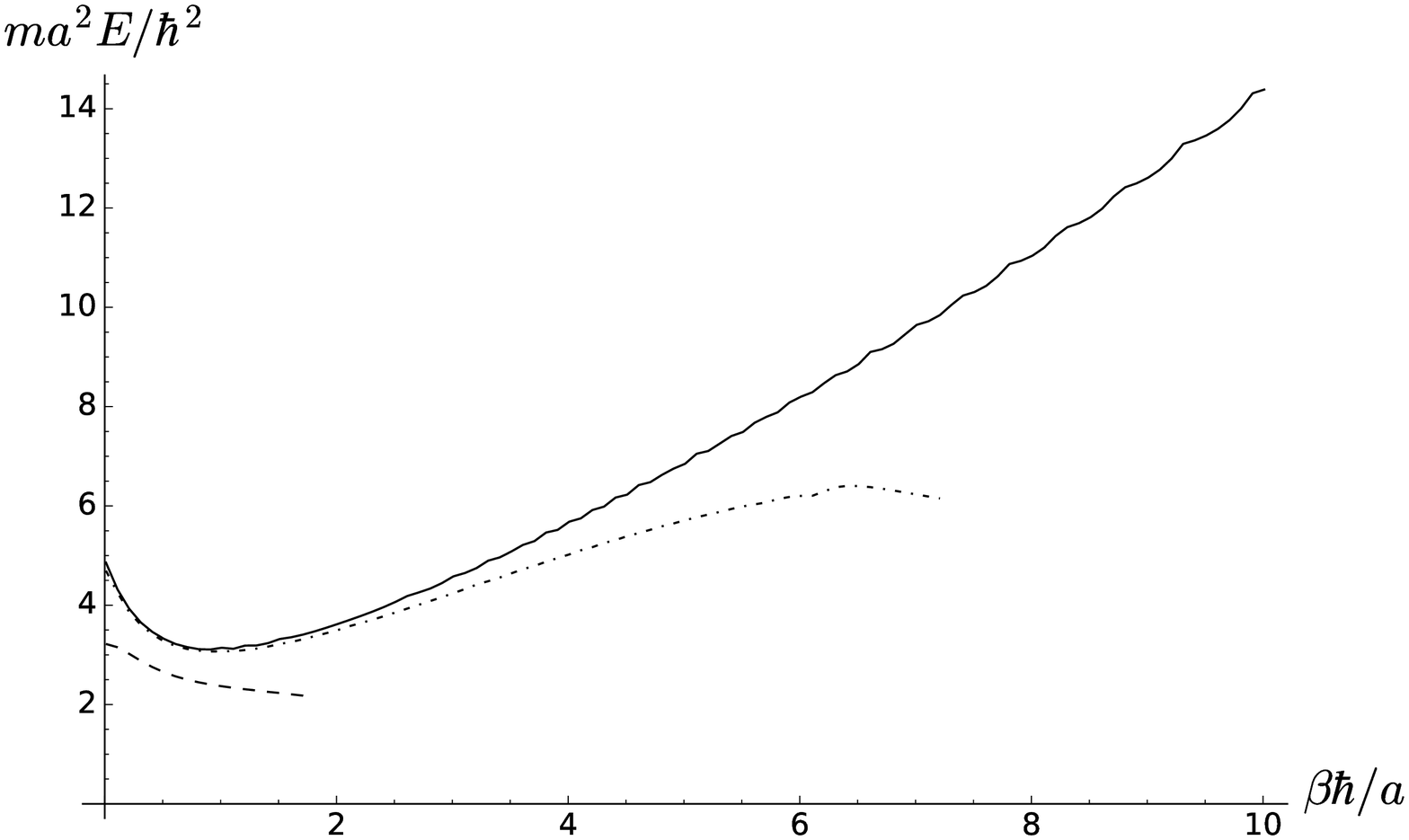}
\includegraphics[width=0.48\linewidth]{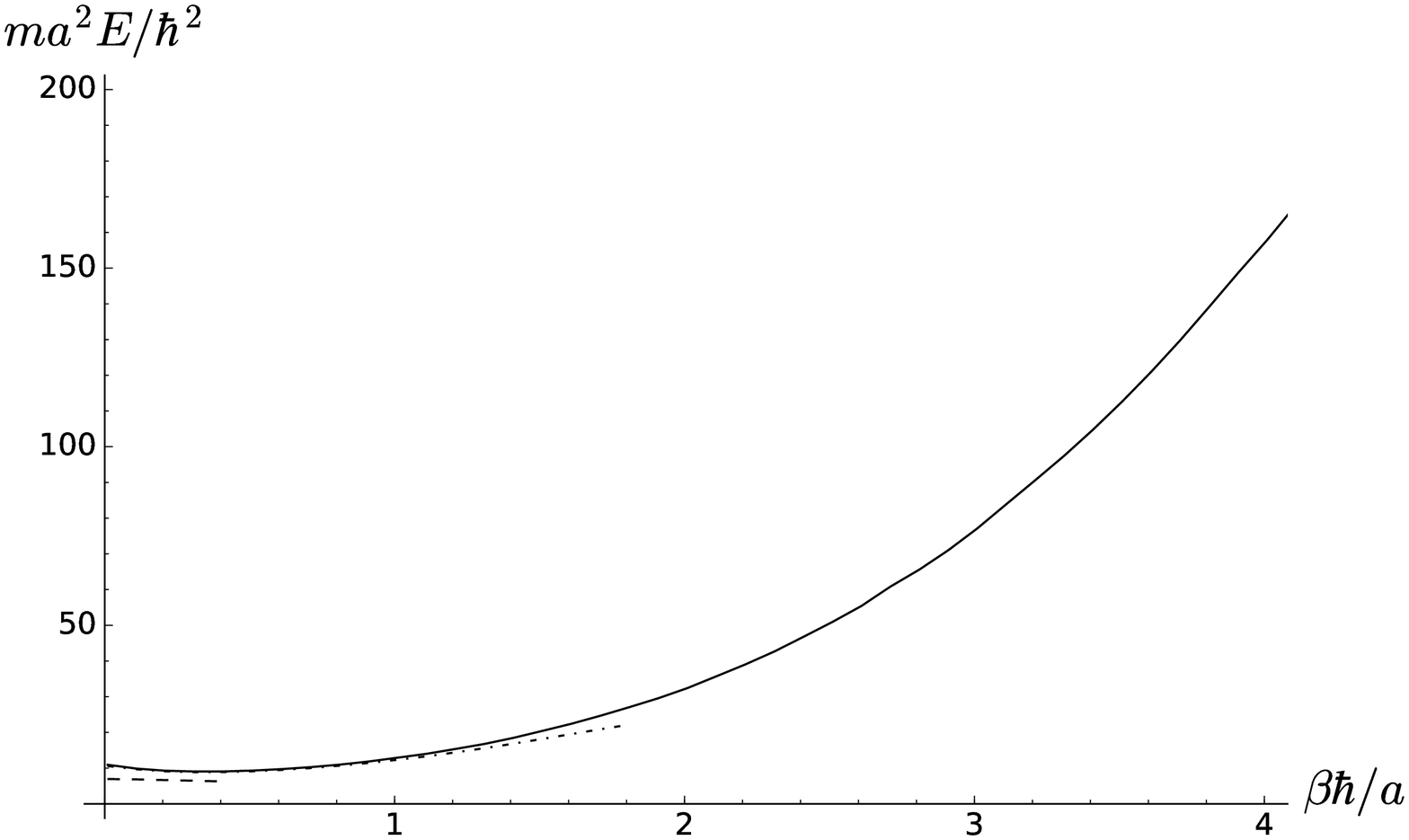}
\includegraphics[width=0.48\linewidth]{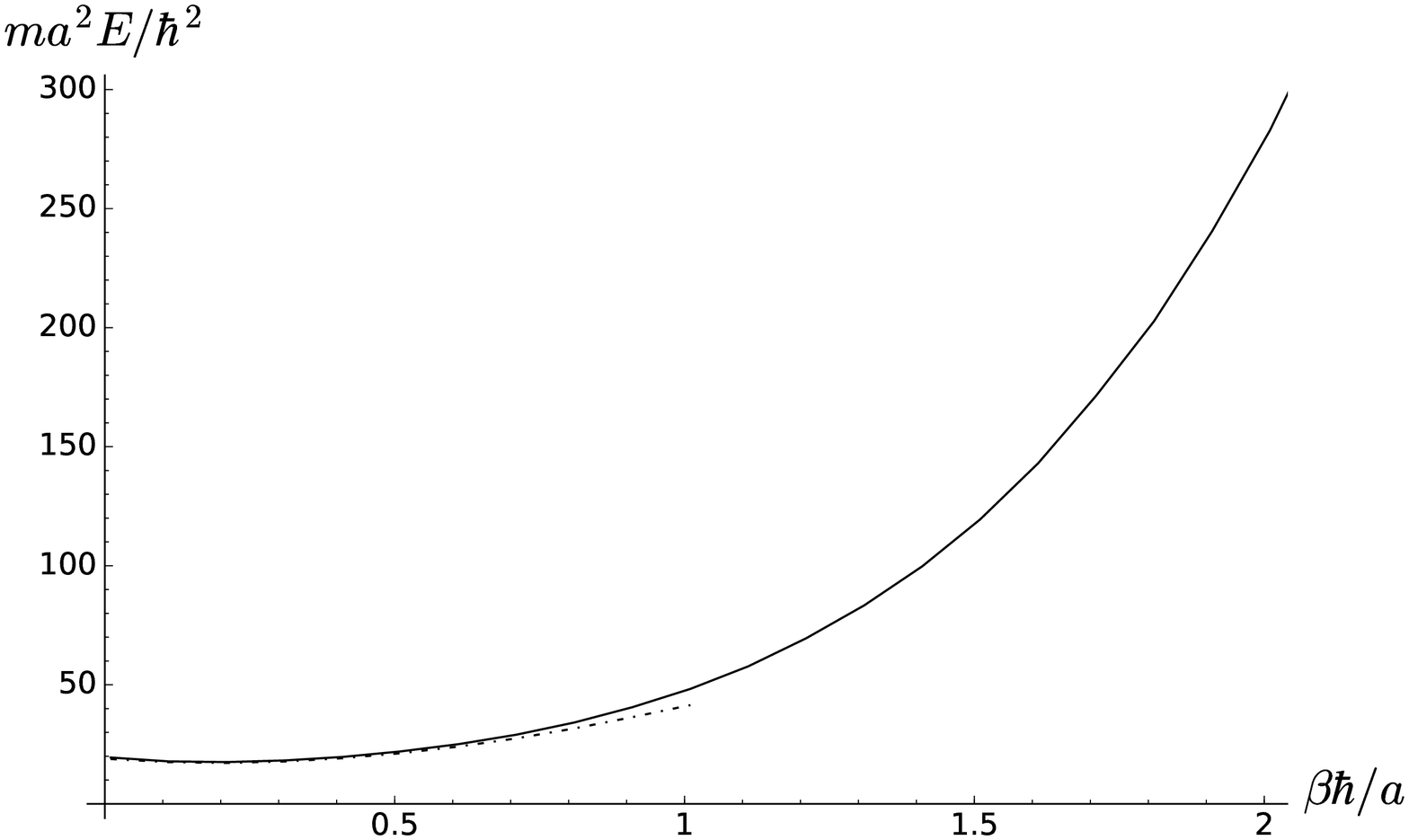}
\caption{Plots of $E$ versus $\b$
for $L=6$ with $ma^2V_0/\hbar^2 = 1 \textrm{ (dotted)}, 10 \textrm{ (dashed)},$
$10^3 \textrm{ (dash-dot)}, 10^6 \textrm{ (solid)}.$
The results are obtained using at least $100$ sample values of $\m.$
From left to right and top to bottom,
the plots are for the ground state, the first excited state,
the second excited state, and the third excited state.}
\label{neqm-ground-fixV-various}
\end{figure}

In Figure~\ref{neqm-ground-fixV-various},
we show plots of $E$ against $\b$
with
$ma^2V_0/\hbar^2 = 1, 10, 10^3, 10^6.$
At each fixed $\b,$
the ground state energy seems to converge
for large $V_0.$ This can be seen in the
figure that there is no noticeable difference
between the cases $V_0 = 10^3\hbar^2/(ma^2)$ (dash-dot line)
and $V_0 = 10^6\hbar^2/(ma^2)$ (solid line).
The first excited state energy
seems to also converge for large $V_0.$
This can be seen in the figure that for each $\b$
from $0$ to around $2a/\hbar,$
there is
only a slight difference of energy
between the cases $V_0 = 10^3\hbar^2/(ma^2)$
and $V_0 = 10^6\hbar^2/(ma^2).$
The qualitative behaviors for other excited states
are similar.
Moreover, the higher the excited state,
the smaller the range of $\b$
in which the energy for
$V_0 = 10^3\hbar^2/(ma^2)$
and $V_0 = 10^6\hbar^2/(ma^2)$
agree.

\begin{table}[h]
\tbl{Coefficients of wavefunctions for each energy level
with $\b = 0.71a/\hbar,$ $V_0 = 200\hbar^2/(ma^2),$
and truncation parameter $L=3.$
Energies for states from the lowest to the highest
are $6.0580\times 10^{-1}\hbar^2/(ma^2),$ $3.0593\hbar^2/(ma^2),$
$9.6365\hbar^2/(ma^2),$ $2.4482\times 10\hbar^2/(ma^2)$.}
{\begin{tabular}{@{}c S[table-format=2.4e+3] S[table-format=2.4e+3] S[table-format=2.4e+3] S[table-format=2.4e+3] @{}} \toprule
 & \multicolumn{1}{c}{ground} & \multicolumn{1}{c}{first excited} & \multicolumn{1}{c}{second excited} & \multicolumn{1}{c}{third excited}\\\colrule
 $A_1\sqrt{a}$ & -3.8022 & -5.8210 & -5.0512 & 1.1259 \\
 $A_2\sqrt{a}$ & -1.6693e3 & -4.4288e3 & -8.0300e3 & 9.8735e3 \\
 $A_3\sqrt{a}$ & -6.7988e5 & -1.8967e6 & -3.8497e6 & 6.2707e6 \\
 $A_4\sqrt{a}$ & -4.3605e7 & -1.2960e8 & -3.0768e8 & 7.4482e8 \\
 $A_5\sqrt{a}$ & 7.7739e1 & 2.3381e2 & 6.5828e2 & -4.3878e3 \\
 $A_6\sqrt{a}$ & 4.6207e4 & 1.4960e5 & 4.6646e5 & -3.2248e6 \\
 $A_7\sqrt{a}$ & 8.8139e6 & 2.8379e7 & 8.5620e7 & -4.9321e8 \\
 $B_1\sqrt{a}$ & -8.0461e-1 & 7.5687e-1 & 6.6383e-1 & 3.8105e-1 \\
 $B_2\sqrt{a}$ & -2.5189e-5 & 2.7911e-5 & 3.0203e-5 & 2.1175e-5 \\
 $B_3\sqrt{a}$ & -3.2560e-10 & 3.8536e-10 & 4.4497e-10 & 3.3552e-10 \\
 $B_4\sqrt{a}$ & -9.1693e-16 & 1.1120e-15 & 1.2284e-15 & 9.1870e-16 \\
 $B_5\sqrt{a}$ & 6.5369e-6 & -1.1918e-5 & -1.5209e-5 & -1.3453e-5 \\
 $B_6\sqrt{a}$ & 4.2656e-11 & -5.9388e-11 & -4.3040e-11 & -3.9307e-11 \\
 $B_7\sqrt{a}$ & -1.2090e-16 & 4.2049e-16 & 9.0814e-16 & 9.0012e-16\\ \botrule
\end{tabular}\label{tab-L3wf} }
\end{table}

After the energy levels are obtained,
one can then proceed to compute wavefunctions for each level.
In Table~\ref{tab-L3wf}, we demonstrated
coefficients for wavefunctions with
$\b = 0.71a/\hbar, V_0 = 200\hbar^2/(ma^2),$
and truncation parameter $L=3.$
In this case, there are four energy levels.
For each level, the wavefunction can be read off
by substituting the coefficients
using corresponding column into Eqs.~\eq{psiI-trunc-new}-\eq{psiIII-trunc-new}
such that for
the ground state and the second excited state,
one sets $B_{e,i} = B_i, B_{o,i} = 0, C_i = A_i; i=1,2,\cdots,7,$
whereas for the first and the third excited states,
one sets $B_{e} = 0, B_{o,i} = B_i, C_i = -A_i; i=1,2,\cdots,7.$

It can be shown by direct integration that
the wavefunctions obtained from Table~\ref{tab-L3wf}
are indeed normalised.
Furthermore, the matching conditions
agree quite well. For example,
in the case of the ground state,
the values of $(\psi_I^{(n)}(-a) - \psi_{III}^{(n)}(-a))a^n\sqrt{a}$
for $n=0,1,2,\cdots,13$
are, respectively,
$9.4792\times 10^{-5}, -1.0214\times 10^{-14}, 4.9794\times 10^{-14}, 5.1514\times 10^{-14}, 1.0258\times 10^{-13}, 6.5654\times 10^{-12}, -1.0525\times 10^{-10}, -1.6844\times 10^{-9}, -6.7354\times 10^{-8}, -1.7243\times 10^{-6}, -3.4485\times 10^{-5}, -9.9325\times 10^{-4}, -3.3547\times 10^{-2}, -6.2158\times 10^{-1}.$
The excited states also share similar behaviors.
That is, the matching conditions are satisfied quite well
for $n\leq 11,$ but then not so well at $n=12, 13.$

\section{Conclusion}\label{sec:Conclude}
In this work, we study NEQM finite square well systems.
The analysis of these systems suggests that
they are differed from the QM counterpart,
though classically they describes the same dynamics.
Furthermore, each of the NEQM systems
are differed from each other.
This is evident from the result that
the energy spectrum and wavefunctions
are dependent on the parameter $\b.$

For each of these systems,
the Schr\"{o}dinger's equation is
an infinite order differential equation,
which leads to the requirement that
the wavefunctions at the well boundary
has to be infinitely differentiable.
In Section \ref{sec:NEQMwell},
we present a way to approximately fulfill the matching
conditions. This is by truncating
the wavefunctions
in such a way that they still satisfy
the NEQM Schr\"odinger's equation
but is differentiable at the well boundary
only up to order $4L+1.$
Our investigation, which is demonstrated
partly in Tables~\ref{tab-sample-gnd}-\ref{tab-sample-fst},
suggests that
the energy spectrum
tend to converge.
In case $\b$ is sufficiently small,
the convergence rate is already good
for small values of $L.$

NEQM finite square well systems
still retain one of the qualitative results
of their QM counterpart.
As $V_0$ decreases,
the bound states vanish one by one,
leaving only the ground state
when $V_0$ is sufficiently small.
This is because the bound state energies
cannot exceed a particular value,
which depends on $V_0.$
For QM finite square well,
the upper bound on the
bound state energy $E$ is $V_0$ itself,
that is $E<V_0.$
However, for NEQM finite square well,
the upper bound on $E$ is now $\b-$dependent,
and is given by
$(\sqrt{1+2m\b^2V_0}-1)/(m\b^2).$

We have also analyzed the behavior for large $V_0.$
We have found that for each fixed $\b,$
the ground state energy tends to converge for large $V_0.$
This is evident in the top left of
Figure~\ref{neqm-ground-fixV-various}.
As for excited states, convergences also seem
to be good especially for sufficiently small $\b.$

It is a well-known fact that the $V_0\to\infty$
limit of QM finite square well coincides with
the QM infinite square well.
So it is interesting to see whether this would
also be the case for the NEQM case.
This question is left as a possible future work.

\section*{Acknowledgements}
We are grateful to Sikarin Yoo-Kong
for bringing the topic of Newton's equivalent Hamiltonians
to our attention
and for various helpful discussions.

\end{document}